\documentclass[aps,twocolumn]{revtex4}
\usepackage[xdvi]{graphicx}

\newcommand{\be}{\begin{equation}}
\newcommand{\ee}{\end{equation}}

\newcommand{\calP}{\mbox{$\mathcal P$}}
\newcommand{\calO}{\mbox{$\mathcal O$}}

\begin{document}
%\draft
\bibliographystyle{prsty}

\title{Improved efficiency with variational Monte Carlo using two level
        sampling}
\author{Mark Dewing \\
        Department of Physics \\
        University of Illinois at Urbana-Champaign, Urbana, Illinois 61801}
 
\email{m-dewin@uiuc.edu}
\affiliation{Department of Physics, University of Illinois at Urbana-Champaign,
  Urbana, Illinois 61801}
\date{\today}

\begin{abstract}
A two level sampling method is applied to variational Monte Carlo (VMC) that
samples the one and two body parts of the wave function separately.
The method is demonstrated on a single Li$_2$ molecule in free space and
32 H$_2$ molecules in a periodic box.
This simple modification increases the efficiency of a VMC simulation
by up to 72\%.

\end{abstract}

\maketitle
\vskip 1cm

Variational Monte Carlo (VMC) \cite{mcmillan65} is an effective 
method for computing 
the ground state properties of atoms, \cite{schmidt90,umrigar88}
molecules,  \cite{filippi96,umrigar99}
and solids. \cite{fahy90,fahy99,williamson96,kralik98}
Explicitly including the electron correlation into the wave function 
allows VMC to recover a large percentage of the correlation energy,
typically 70-90\% (or more for small atoms and molecules). 
\cite{schmidt90,filippi96,lin00}

VMC computes matrix elements with respect to the square of a trial 
wave function, $\psi$.
The most common matrix element is the energy,
\be \label{vmc}
 E = \frac{\int dR\ \psi^2(R;a) E_L(R;a)}{\int dR\ \psi^2(R;a)}
\ee
where $E_L = \frac{1}{\psi}H \psi$.
The variational principle states that this energy will be greater than
or equal to
the true ground state energy for any $\psi$.   Typically, the wave function
is parameterized and those parameters are varied until a minimum  energy
\cite{lin00}
(or alternatively, minimum variance \cite{schmidt90,williamson96}) is reached. 

The Metropolis method \cite{metropolis53}
constructs a transition probability, $\calP(s' \rightarrow s)$,
for use in a Markov process. The result of that process is
the desired normalized probability distribution.
A correct transition probability can be constructed by satisfying detailed 
balance,
\be \label{db}
\pi(s) \calP(s \rightarrow s') =  \pi(s') \calP(s' \rightarrow s)
\ee
where $\pi(s)$ is the desired probability distribution 
($\psi^2(R)$ for VMC ).
In the generalized Metropolis method, the transition
probability is split into two pieces, an {\it a priori} 
sampling distribution, $T(s\rightarrow s')$ 
and an acceptance probability $A(s \rightarrow s')$.

The original Metropolis method generates a trial move, $R'$,
of one particle inside a box of side $\Delta$ centered about the old 
position, $R$. 
This leads to a constant $T$ inside the box and zero outside.
The trial move
is then accepted or rejected with probability
\be \label{accept}
 A = \min \left[ 1, \frac{\psi^2(R')}{\psi^2(R)} \right]
\ee

In electronic problems, a typical wave function is
\be
 \psi = D e^{-U}
\ee
where $D$ is the product of spin up and spin down  Slater determinants of 
single particle
orbitals and $U$ is the two (or higher) body Jastrow factor. This latter
part contains the explicit electron-electron correlation.

A simulation step consists of a trial move of every electron.
Each trial move is accepted or rejected according to  
Eq (\ref{accept}).

A single electron is moved at a time, which only changes one column of the
Slater matrix.  By saving and using the inverse of this matrix, computing
the new determinant and its derivatives is an $\calO(N)$ operation.
\cite{ceperley77}
If the move is accepted, the inverse matrix can be updated with $\calO (N^2)$
operations.
(Note that computing the determinant directly is of order $\calO(N^3)$.)
A smaller acceptance ratio will be faster, since fewer updates need to be
performed.

Multilevel sampling has been used extensively in path integral Monte Carlo.
\cite{ceperley95}   
In multilevel sampling, the wave function is split into several factors (levels),
and 
an accept/reject step is performed after computing each factor.
The entire wave function is computed only if there is an acceptance after
every factor.   Rejections will make the overall algorithm faster,
since not all the factors need to be computed.
For VMC it is  natural to use two factors -  the single body part,
$D$, as one factor and the
two body part, $e^{-U}$, as the other.
The single body part is the cheaper
of the two to compute,  so computing it first will give the greatest
reduction in time.

The two level sampling algorithm for VMC proceeds as follows.
A trial move, $R'$, is proposed and
accepted at the first stage with probability
\be
 A_1 = \min \left[ 1, \frac{D^2(R')}{D^2(R)} \right]
\ee

If accepted at the first stage, the two body part is computed and the trial 
move is accepted with probability
\be
A_2 = \min \left[ 1, \frac{\exp\left[-2U(R')\right]}{\exp\left[-2U(R)
\right]} \right]
\ee
It can be verified by substitution that these satisfy detailed balance in 
Eq. (\ref{db}).  
After an acceptance at this second level, the inverse Slater matrices
are updated as described previously.

The figure of merit for these simulations is the efficiency
\be
\xi =  \frac{1}{\sigma^2 T}
\ee
where $T$ is the computer time and $\sigma$ is the error estimate
of the average of some property such as the energy.

Maximum efficiency results from a competition between two effects.
The first is that a lower acceptance ratio will yield a lower run time and
a larger efficiency.
The second is serial correlations.  A lower acceptance ratio will give
a longer correlation time, hence a larger error and lower efficiency.

In maximizing the efficiency, the obvious parameter to adjust is the step size,
$\Delta$.   But the number of steps between
computations of $E_L$ can also be adjusted.
The Metropolis algorithm produces correlated state points, so successive
samples of $E_L$ don't contain much new information.
It is advantageous to compute $E_L$  every few steps
rather than every step.  In this work the local energy was computed
every 5 steps.

The algorithm is demonstrated using
a Li$_2$ molecule in free space and a collection of 32 H$_2$ molecules in a 
periodic box.
The wave functions, which are the $\Psi_{\mathrm{III}}$'s from  Reynolds, 
{\it et al.},
\cite{reynolds82} 
use a simple electron-electron and electron-nuclear Jastrow term, and use
floating Gaussians for the orbitals.
The hydrogen molecules were in box of side 19.344 atomic units ($r_s = 3.0$).

The results for the different sampling methods for the Li$_2$ molecule 
are given in Tables \ref{li2regular} and \ref{li2twolevel}.
The second level acceptance ratio
is quite high, indicating the single body part is a good approximation
to the whole wave function.
The efficiency is also shown in Figure 1.

The results for the different sampling methods for the H$_2$ molecules
are given in Tables \ref{h32regular} and \ref{h32twolevel}.
The efficiency is also shown in Figure 2.
%Note the efficiency curve has two maxima.

Comparing the maximum efficiency for each sampling method, two level
sampling is 39\% more efficient than the standard sampling for Li$_2$, 
and 72\% more efficient for 32 H$_2$'s.

More complicated schemes using more levels or different splittings of the
wave function could be devised.
This particular scheme is attractive because it uses quantities 
readily available in a VMC computation, and requires minimal modification
to existing VMC sampling algorithms.

%\section{Acknowledgments}
This work has been supported by the computational facilities at NCSA and
by NSF grant DMR 98-02373.

%------------   Li 2
\begin{table}\caption{Timings for Li$_2$ molecule using
the standard sampling method. All times in seconds on an SGI Origin 2000.}
\label{li2regular}
\begin{tabular}{cccccc}
               &  Acceptance &   Determinant   & Jastrow  & Total  &\\ 
$\Delta$       &   Ratio     &     Time        &   Time   &  Time & $\xi$\\
 \hline
 1.0    & 0.610   & 48.3  & 340  & 516 & 1190 \\
 1.5    & 0.491   & 48.1  & 340  & 508 & 1680 \\
 2.0    & 0.407   & 48.2  & 340  & 503 & 1460 \\
 2.5    & 0.349   & 48.2  & 339  & 499 & 1070 \\
 3.0    & 0.307   & 48.2  & 339  & 496 &  800 \\
\end{tabular}
\end{table}

\begin{table}\caption{Timings for Li$_2$ molecule using
the two level sampling method. All times in seconds on an SGI Origin 2000.}
\label {li2twolevel}
\begin{tabular}{cccccc}
    &  First Level     &  Second Level  & Total Acc. &  \\
$\Delta$  & Acc. Ratio & Acc. Ratio & Ratio &Time &  $\xi$ \\ 
 \hline
1.0       & 0.674   & 0.899   & 0.606  & 400  & 1580 \\
1.5       & 0.543   & 0.894   & 0.485  & 347  & 2430 \\
2.0       & 0.447   & 0.897   & 0.401  & 304  & 2340 \\
2.5       & 0.379   & 0.902   & 0.342  & 276  & 1910 \\
3.0       & 0.331   & 0.906   & 0.300  & 256  & 1400 \\
\end{tabular}
\end{table}

%------------   H_64

\begin{table}\caption{Timings for 32 H$_2$ molecules in a periodic box using
the standard sampling method. All times in seconds on a Sun Ultra 5.}
\label{h32regular}
\begin{tabular}{cccccc}
               &  Acceptance &   Determinant   & Jastrow  & Total  &\\ 
$\Delta$       &   Ratio     &     Time        &   Time   &  Time  & $\xi$ \\ 
  \hline
 2.0    & 0.606   & 167  & 1089  & 2015 & 0.61 \\
 3.0    & 0.455   & 167  & 1085  & 1891 & 1.22 \\
 4.0    & 0.338   & 166  & 1084  & 1794 & 1.23 \\ 
 5.0    & 0.250   & 166  & 1080  & 1722 & 1.06 \\
 6.0    & 0.185   & 164  & 1080  & 1668 & 1.02 \\
 7.0    & 0.139   & 162  & 1084  & 1629 & 0.76 \\
\end{tabular}
\end{table}

\begin{table}\caption{Timings for 32 H$_2$ molecules in a periodic box using
the two level sampling method. All times in seconds on a Sun Ultra 5.}
\label {h32twolevel}
\begin{tabular}{cccccc}
    &  First Level     &  Second Level  & Total Acc. & Total & \\
$\Delta$  & Acc. Ratio & Acc. Ratio & Ratio &Time &  $\xi$ \\ 
 \hline
2.0       & 0.740   & 0.795   & 0.589  & 1804  & 0.59 \\
3.0       & 0.598   & 0.728   & 0.436  & 1421  & 1.77 \\
4.0       & 0.468   & 0.681   & 0.319  & 1185  & 2.11 \\
5.0       & 0.357   & 0.649   & 0.232  &  994  & 1.55 \\
6.0       & 0.370   & 0.627   & 0.169  &  849  & 1.87 \\
7.0       & 0.204   & 0.609   & 0.124  &  740  & 1.46 \\
\end{tabular}

\end{table}

% Figure 1
\begin{figure}
\begin{center}
   \includegraphics[width=8.5cm]{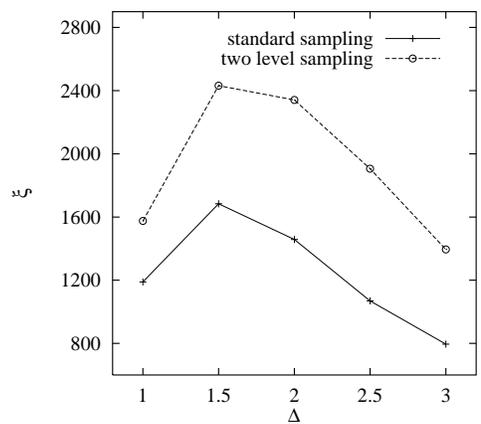} 
   \caption{ Efficiency of VMC for Li$_2$.}
\end{center}
\end{figure}

% Figure 2
\begin{figure}
   \includegraphics[width=8.5cm]{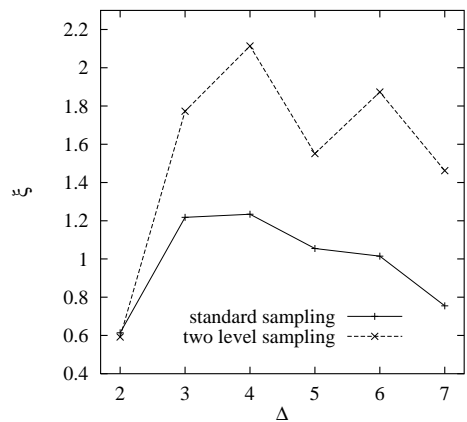} 
   \caption{ Efficiency of VMC for 32 H$_2$ molecules.}
\end{figure}

\end{document}